\documentclass[prb,aps,twocolumn,amsmath,amssymb,floatfix,
superscriptaddress]{revtex4}

\usepackage[dvips]{graphics}
\usepackage{color}
\definecolor{dred}{rgb}{0,0,0.6}

\begin{document}

\title{Charge-based re-programmable logic device with built-in memory:
New era in molecular electronics}

\author{Moumita Patra}

\affiliation{Physics and Applied Mathematics Unit, Indian Statistical
Institute, 203 Barrackpore Trunk Road, Kolkata-700 108, India}

\author{Santanu K. Maiti}

\email{santanu.maiti@isical.ac.in}

\affiliation{Physics and Applied Mathematics Unit, Indian Statistical
Institute, 203 Barrackpore Trunk Road, Kolkata-700 108, India}

\begin{abstract}

We put forward a new proposal of designing charge-based logic devices 
considering a cyclic molecule that can be programmed and re-programmed 
for different functional logical operations and suitably engineered for 
data storage as well. The key idea is based on the appearance of bias 
induced circular current under asymmetric molecule-to-electrode interface
configuration which does not dissipate even when the bias is off. Our 
results are valid for a broad range of parameter values, and provide 
a boost in the field of storage mechanism, reconfigurable computing, 
charge-based logic functions and other nano-scale applications.

\end{abstract}

\maketitle

\section{Introduction}

Designing of logic gates at nano-scale level based on molecules has been the 
subject of growing attention as they are treated as basic building blocks 
of digital nanoelectronics~\cite{mlg1,mlg2,mlg3,mlg4,mlg5,mlg6,smnew1}. 
Among different molecular systems, cyclic molecules play the central role 
because of their widespread applications in electronic devices due to high 
integration density, low cost and chemical stability. Following the proposal 
of molecular logic gates by de Silva and his group~\cite{mlg1}, interest 
in this subject has rapidly picked up, circumventing the use of FETs and 
MOSFETs. Instead of designing individual logic gates it is always beneficial 
to construct functional programmable logic devices 
(PLDs)~\cite{pld1,pld2,pld3,pld4} and it becomes more versatile if storage 
mechanism can also be implemented as well. The most significant
advantage of a PLD is that several logic functions can be programmed and
re-programmed from a single device yielding greater performance, reducing
built-in elements and many other suitable functional operations. In addition
to this, the computing performance can be improved enormously if PLDs
are capable of storing output as in conventional logic systems information 
need to be transferred in a storage device for preventing them getting lost 
as they are volatile~\cite{pld2,pld3,pld4}. {\em Though few proposals are 
available for the description of logical operations with built-in-memory 
based on spin degree of freedom~\cite{pld2,pld3,pld4}, but no attempt has 
been made so far for devising a charge based PLD that can also store 
computable information, and in this communication, we essentially focus 
along this direction.} 
No doubt, spin-based systems have several advantages~\cite{mr1}, however 
they are not suitable enough to hit the present market~\cite{stt3}. 
The main concern is the proper regulation of spin states which is 
still not yet clear and needs further probing. Whereas for charge-based 
systems this issue is not involved any more.

In this communication we explore how charge-based logic gates can be designed 
from a single benzene molecule that can be used as a storage device as well.
Five logical operations (OR, NOT, XOR, XNOR, NAND) are designed from the 
benzene ring, though other two (AND, NOR) can eventually be performed from 
the rest. Very recently, in a separate work, we have described how three 
primary logic gates, OR, AND and NOT, can be implemented considering a 
benzene molecule exploiting the effect of quantum interference among the 
electronic waves passing through different arms of the molecular 
ring~\cite{smnew1}. The essential mechanism was based on the appearance 
of anti-resonant states under a specific molecule-to-electrode interface 
geometry, and no concept of bias induced circular current and storage 
mechanisms have been put forward. 
\begin{figure}[ht]
{\centering\resizebox*{8.5cm}{5.35cm}{\includegraphics{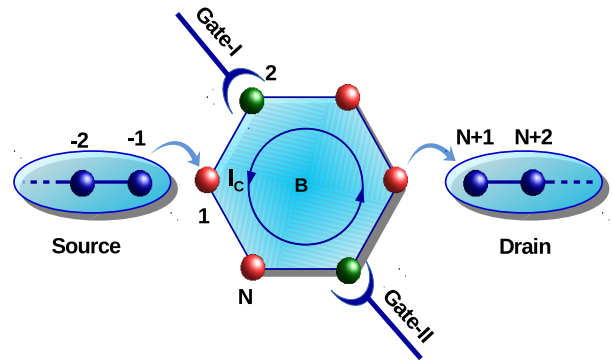}}\par}
\caption{(Color Online). Model molecular junction where a benzene ring is
coupled to source and drain electrodes in such a way that the upper and lower
arms acquire equal lengths. Two gate electrodes (gate-I and Gate-II) are used
to tune site energies of the filled green colored sites those are considered
as inputs of the logic functions. The other sites (filled red circles), not
influenced by external gate electrodes, are the parent lattice sites. By
breaking the symmetry between the two molecular arms a net circular current
($I_c$) is established which, on the other hand, induces a finite magnetic
field $B$. The complete system is simulated by the tight-binding framework.}
\label{fig1}
\end{figure}
The prescription of logical operations that we are going to discuss in this 
communication is completely new, to the best of our knowledge, and a 
successful conclusion of logic operations along with storage mechanisms 
(charge-based) will definitely boost the field of digital molecular 
electronics. The central idea of the present work is that a net circular 
current (charge) is established 
in the molecular ring by suitably connecting it with two electron baths 
(source and drain)~\cite{cir1,cir2,cir3}, and once the current is established 
it persists even when the bias is off, analogous to persistent current in an 
isolated conducting Aharonov-Bohm ring~\cite{pc1,pc2,pc3,pc4,pc5} though 
the origin is quite different. In presence of finite bias, circular current 
in molecular ring appears only when the symmetry between two arms of the 
junction is lost. 
This can be done in two ways, either by making unequal lengths of upper 
and lower arms of the molecular junction or by introducing an asymmetry in 
anyone of the two arms for a lengthwise symmetric molecular junction. 
We follow the second step where a benzene molecule is coupled 
symmetrically to source and drain and applying suitable gate voltages, 
associated with inputs of logic functions, symmetry gets broken to yield 
non-zero circular current ($I_c$). A finite $I_c$ corresponds to ON state 
of the output, whereas $I_c=0$ represents the OFF state. When a gate electrode, 
placed in the vicinity of any carbon site among six sites, is ON its site
potential gets modified~\cite{baer1,baer2} resulting an asymmetry. Thus 
placing gate electrode(s) in appropriate locations possible logic functions 
can be configured from the single molecular junction. And since the current 
persists even in the absence of bias, the device can be used for storage 
purpose as well. Depending on $I_c$ (finite or zero) it can store $1$ or $0$ 
logic bits.

The work is arranged as follows. In Sec. II we describe a general model of
molecular junction comprising a phenyl ring which is the key element of our 
system and illustrate the theoretical prescription for the calculation of 
circular current under different conditions. The essential results which 
include logic functions and storage mechanisms are given in Sec. III. 
Finally, we summarize our key findings in Sec. IV.

\section{Molecular model and Theoretical prescription}

Let us begin with the model molecular junction shown in Fig.~\ref{fig1} 
where a benzene
molecule is sandwiched between source (S) and drain (D) electrodes. These 
electrodes are assumed to be perfect, one-dimensional and semi-infinite, and 
they are coupled to the phenyl ring symmetrically (upper and lower arms have 
equal lengths) to form a lengthwise symmetric molecular junction. As already
mentioned that to get circular current we need to break symmetry between the 
two arms for this lengthwise symmetric junction which we do by altering the
status of a single atomic site or more sites depending upon the requirement
of logic functions. Two different status are enumerated by two filled colored
sites (green and red), where the filled red circles correspond to the parent 
lattices having a specific site energy and the other circles (i.e., green 
circles) have another site energy, and this change of site energy is done by
means of gate electrode. 

The full system is modeled by a tight-binding (TB) framework and the 
Hamiltonian becomes 
\begin{equation}
H=H_M + H_S + H_D + H_{tun}
\label{eqn1}
\end{equation} 
where $H_M$ represents the molecular Hamiltonian, $H_S$ and $H_D$ are the 
Hamiltonians for the S and D electrodes, and the Hamiltonian $H_{tun}$ is 
associated with the coupling of the molecule with side-attached electrodes.
All these Hamiltonians are described by a similar kind of TB form and within
non-interacting picture the general form of Hamiltonian reads as~\cite{cir1,cir3}
\begin{equation}
H_K=\sum_p \epsilon_p c_p^{\dagger} c_p+\eta\sum_p \left(c_{p+1}^{\dagger}c_p
+ c_p^{\dagger}c_{p+1}\right)
\label{eqn2}
\end{equation}
where $K=M$, $S$, $D$ and $tun$ as used in Eq.~\ref{eqn1}. The parameters 
$\epsilon_p$ and $\eta$ describe the site-energy and nearest-neighbor hopping 
(NNH) integral, respectively, and $c_p^{\dagger}$, $c_p$ are the Fermionic 
operators. Now, depending on different parts of the molecular junction we 
use different symbols of site energy and NNH integrals. For the two perfect 
electrodes these are $\epsilon_0$ and $t_0$, respectively, while for the 
molecule, having six atomic sites, these parameters are $\epsilon_m$ and $v$, 
respectively. The molecular ring is coupled to S and D through the 
coupling parameter $\tau_S$ and $\tau_D$, respectively, and these are 
responsible for the tunneling of electrons between the molecule and 
electrodes.

This is all about the Hamiltonian of our system. Now, in order to explore 
logical operations and storage mechanisms first we need to calculate bias 
induced circular current, and we do it by using wave-guide 
theory~\cite{cir1,cir2,cir3,wg1,wg2,wg3}, a standard method of studying 
electron transport through a conducting junction. In this prescription a 
set of coupled linear equations involving wave amplitudes at different 
lattice sites of the molecule are solved those are generated from the 
Schr\"{o}dinger equation
\begin{equation}
H|\psi\rangle=E|\psi\rangle
\label{eqn3}
\end{equation}
where $|\psi\rangle=\sum_p C_p|p\rangle$. $C_p$'s are the wave amplitudes and 
$|p\rangle$'s are the Wannier states. Assuming plane wave incidence from the 
source electrode with unit amplitude we can write the wave amplitude at any 
particular site $q$ (say) of S as $e^{ik(q+1)a}+re^{-ik(q+1)a}$, where $k$ 
is the wavevector and it is determined from the relation 
$E=\epsilon_0 + 2 t_0 \cos(ka)$ ($a$ being the lattice spacing). $r$ is the 
reflection coefficient. For the drain electrode the wave amplitude at any 
particular site $l$ (say) becomes quite simpler as there is no reflection 
from the drain end and gets the form $te^{ikla}$, where $t$ is the 
transmission coefficient, and, its absolute square gives the transmission 
probability. Using the above wave forms of incident and transmitted waves,
we solve the coupled linear equations involving wave amplitudes at different 
lattice sites for different injecting electron energies, and then we 
calculate bond current density between any two neighboring sites 
(say, $m$ and $m+1$) of the molecule following the 
expression~\cite{cir1,cir3,wg1}
\begin{equation}
J_{m,m+1}=\frac{2e}{\hbar} \mbox{Im} \left[v C_m^* C_{m+1}\right]
\label{eqn4}
\end{equation}
Integrating this current density we compute bond current for a particular bias
voltage $V$ and it becomes~\cite{cir1,cir3,datta}
\begin{equation}
I_{m,m+1}(V) = \int\limits_{E_F-\frac{eV}{2}}^{E_F+\frac{eV}{2}}
J_{m,m+1}(E) \, dE
\label{eqn5}
\end{equation}
where $E_F$ is the equilibrium Fermi energy. Once the bond current is 
evaluated, the net circular current in the molecular ring is obtained from 
the relation~\cite{cir1,cir3,sm1,sm2}
\begin{equation}
I_c=\frac{1}{L} \sum_i I_{m,m+1}\, a
\label{eqn6}
\end{equation}
where $L$ ($=Na$) describes the circumference of the molecular ring geometry 
and $N$ corresponds to the total number of atomic sites in it. The circular 
charge current can be positive or negative as well depending on bias regimes, 
unlike transport charge current which is always positive. We assign positive 
sign for the
circular current flowing in the counter-clockwise direction. Thus, for a 
lengthwise symmetric molecular junction when the status of the upper and lower
arms are exactly identical, equal currents pass through these arms but they 
propagate in opposite directions which leads to a vanishing circular current. 
We refer it as OFF state for the logical operations. On the other hand, for 
the asymmetric conditions currents in different arms are no longer identical 
which results a finite $I_c$ and we define it as ON state of the output. 
The non-zero circular current produces a magnetic field $B$, and we can 
easily calculate it ($B$) at any desired point $\vec{r}$ by using 
Biot-Savart's law~\cite{cir1,cir2,cir3,sm1,sm2}
\begin{equation}
\vec{B}(\vec{r}) = \sum\limits_{\langle m,n \rangle} \left(\frac{\mu_0}{4\pi}
\right)
\int I_{m,n}\frac{d\vec{r^{\prime}} \times(\vec{r}-\vec{r^{\prime}})}
{|\vec{r}-\vec{r^{\prime}})|^3}
\label{bb}
\end{equation}
where $r^{\prime}$ is the position vector of the current element 
$I_{m,n} d\vec{r^{\prime}}$, and $\mu_0$ ($=4\pi \times 10^{-7}$ NA$^{-2}$) 
is the magnetic constant. To calculate magnetic field we need to take the
sum over all bonds $\langle m,n\rangle$ of the molecular ring. This 
circular current induced magnetic field may be utilized for storage 
mechanism as well, as we discuss below.

For the entire discussion we set the system temperature at absolute zero,
for simplification, and no physics will be altered at all at finite (low) 
temperatures since the system size is too small which yields much higher 
average spacing of energy levels and at the same time the thermal energy 
broadening is extremely small compared to the broadening caused by the 
molecule-to-electrode coupling~\cite{datta}.

\section{Numerical Results and Discussion}

In what follows we present our results. Before scrutinizing the individual logical
operations along with storage functions, based on the mechanism of bias induced 
circular current under two input conditions, we explore the general features 
of circular current $I_c$ in the molecular ring. In order to explore it, let 
\begin{figure}[ht]
{\centering\resizebox*{8.5cm}{10cm}{\includegraphics{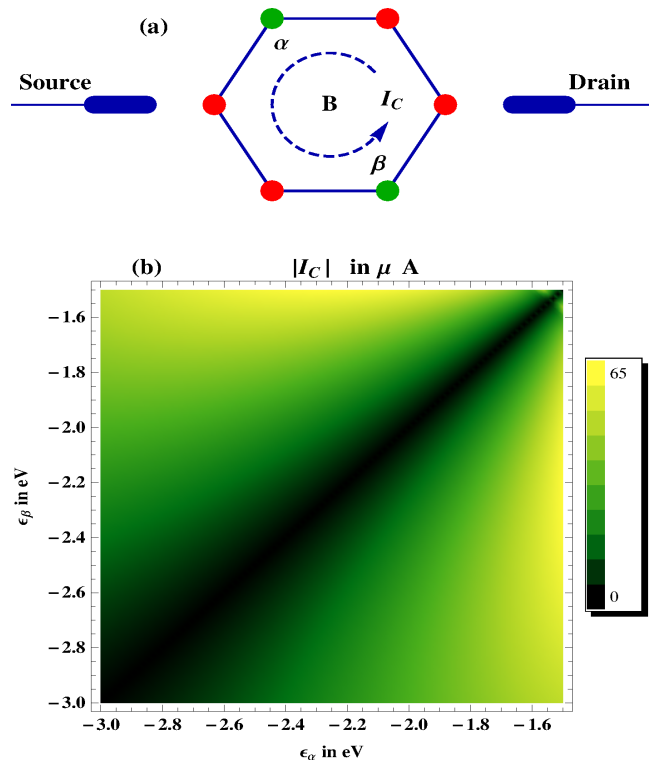}}\par}
\caption{(Color Online). Molecular setup, shown in (a), along with the 
dependence of circular current, shown in (b) (density plot), on 
$\epsilon_{\alpha}$ and 
$\epsilon_{\beta}$ for this setup. Here we plot the absolute values of the 
circular current (though it can have both positive and negative values, 
unlike the conventional transport current which is always positive) since 
the output of our logic functions depends on whether the circular current 
is finite (ON state) or zero (OFF state). For numerical calculations we choose
the site energy and nearest-neighbor hopping (NNH) strength in S and D as zero
and $3\,$eV, respectively. The site energies of the parent lattice sites 
(filled red circles) of the molecular ring are fixed at $-1.5\,$eV, and for 
the other two sites (filled green circles) site energies are variable. 
The NNH integral in the molecular ring is set at $2.5\,$eV, and this 
molecule is coupled to S and D with the hopping strength $1\,$eV. All the
results are computed at the bias voltage $V=2\,$Volts.}
\label{Circular}
\end{figure}
us start with the setup given in Fig.~\ref{Circular}(a) where the input 
signals are applied to the filled green circles marked as $\alpha$ and 
$\beta$ sites. These two sites get two different energies, referred as 
$\epsilon_{\alpha}$ and $\epsilon_{\beta}$, compared to the parent lattice 
sites (shown by the filled red circles) whose site energies are parameterized 
by $\epsilon_m$ as mentioned earlier in theoretical prescription. For this 
typical junction setup (Fig.~\ref{Circular}(a)), the dependence of circular 
current on $\epsilon_{\alpha}$ and $\epsilon_{\beta}$ is shown in the lower 
panel of Fig.~\ref{Circular} (Fig.~\ref{Circular}(b)).
\begin{figure*}[ht]
{\centering\resizebox*{12cm}{12cm}{\includegraphics{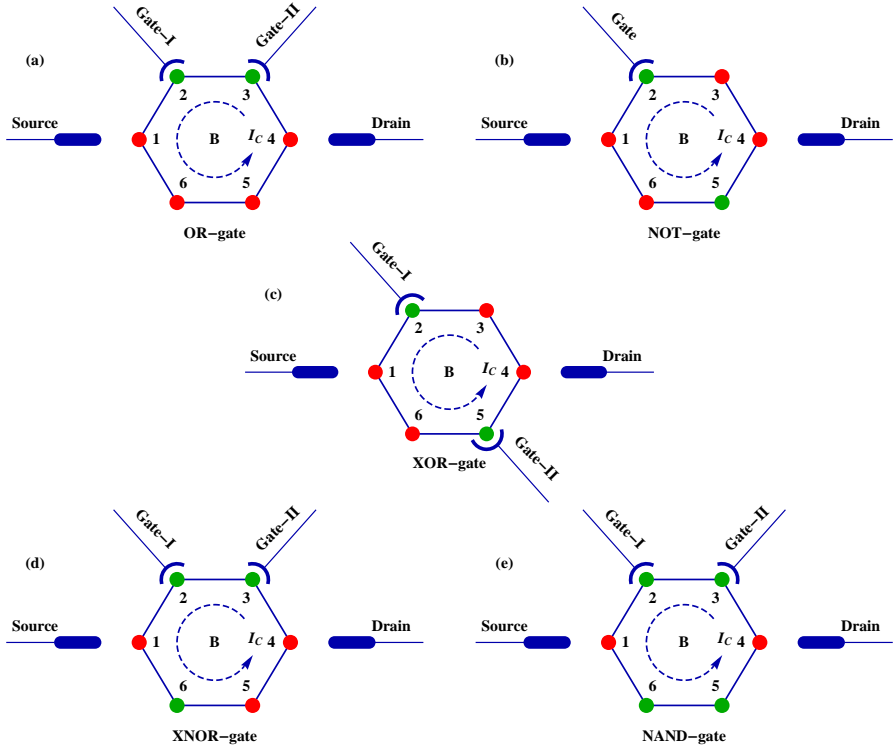}}\par}
\caption{(Color Online). Layouts of molecular junctions to get distinct 
logical operations. Depending on the operations we vary the number of 
filled green circles as well as their locations on the ring circumference. 
When the gate electrode is ON, the green site acquires the energy $-3\,$eV
and we consider it as ON state of the input, while for the OFF state of
the input (viz, when gate is OFF) this energy becomes identical with the 
parent lattice sites (filled red circles) which is fixed at $-1.5\,$eV.
The additional filled green circles, where gate electrodes are not placed
in the setups, will have the fixed site energy (equals to $-3\,$eV) for 
the entire process. The reason behind this consideration is explained 
clearly in the text. All the other physical parameters are same as taken 
in Fig.~\ref{Circular}. Total five logical operations are presented.}
\label{f1}
\end{figure*}
We tune $\epsilon_{\alpha}$ and $\epsilon_{\beta}$ in a wide range starting 
from $-3\,$eV to $-1.5\,$eV, and interestingly we find that the circular current
completely vanishes whenever $\epsilon_{\alpha}$ matches with $\epsilon_{\beta}$
for this lengthwise symmetric molecular junction. Whereas, finite $I_c$ is 
observed under the condition $\epsilon_{\alpha}\ne\epsilon_{\beta}$, and 
it ($I_c$) reaches
to a maximum when the difference between these two inputs are maximum. The
above facts can be explained clearly by incorporating the concept of symmetry 
between the upper and lower arms of the molecular junction. When 
$\epsilon_{\alpha}=\epsilon_{\beta}$, both these two arms are physically 
identical, and therefore, the currents passing through these arms are equal
in magnitude but as they traverse in the opposite directions the net circular
current vanishes. Naturally, a finite current appears only when the symmetry
between the two arms is broken, and that is done by making the site energies
of the two green sites unequal. The underlying physics of getting circular
current (resultant of individual bond currents, see Eq.~\ref{eqn6}) is that
at least one molecular energy level has to appear within the bias window 
for which we are interested to get circular current. With increasing this
voltage window, more energy levels may appear and all of them contribute 
to the current (both positive and negative currents are highly expected). 
Eventually the net response is determined by the dominating energy levels. 
So, naturally, we can expect both the two signs of this current, unlike 
the transport charge current which always be positive. As for the logical 
operations we are not bothering about the sign of the current $I_c$, in 
Fig.~\ref{Circular}(b) we take the absolute values of $I_c$ to have a
clear idea of this density plot. The other notable thing is that the 
amount of circular current is too high (few microamperes) which induces
a strong enough magnetic field ($\sim$milliTesla) at and away (not too 
far) from the ring center.

This is the basic concept of circular current that is utilized to explore 
different logical operations, and below we characterize them. In this 
communication we are able to explore total five logic gates (OR, NOT, 
XOR, XNOR, NAND) based on the idea of bias induced circular current and 
the sketches for designing these gates are shown in Figs.~\ref{f1}(a) 
to (e), respectively. Two gate electrodes (Gate-I and Gate-II) are used 
for two-input logic gates, while a single such electrode is used for 
the one-input logic gate (i.e., for NOT gate). Applying a suitable 
gate voltage we tune the site energies ($\epsilon_{\alpha}$, 
$\epsilon_{\beta}$) 
of the atomic sites placed in the vicinity of gate electrodes. When the 
gate is ON, the site energy becomes $-3\,$eV, while in the OFF state this
energy becomes identical to the parent lattice sites which is fixed at 
$-1.5\,$eV. These two site energies are considered as the ON and OFF states 
of an input signal. Now, in addition to this, in some cases (NOT, XNOR, 
and NAND) we include green site(s) whose energy is fixed at $-3\,$eV for 
the entire operation. To set this energy we also use an external gate 
electrode, like other cases, but as it is not changed we do not show the 
image of extra gate electrodes in our schematic diagrams, not to make 
them clumsy. These additional green sites (for the NOT and XNOR gates the 
number of extra green sites is one, while for the XNOR gate this number 
is two) are required to achieve definite logical operations following the 
symmetry conditions. With these propositions now the logical operations 
can be clearly understood from the distinct sketches shown in 
Fig.~\ref{f1}(a)-(e). Their output responses are presented in 
Table~\ref{tab1} for the quantitative treatment. Looking into the data 
given in Table~\ref{tab1} we see that, in each case quite large circular 
\begin{table}[ht]
\caption{Truth tables for different logical operations. High circular current
(in units of $\mu$A) is obtained in each case.}
$~$
\vskip -0.25cm
\fontsize{7}{11}
\begin{tabular}{|c|c|c|c|c|c||c|c|}
\hline
\textbf{Input} & \textbf{Input} &
\multicolumn{4}{c||}{\textbf{Output}} &
\multicolumn{2}{c|}{\textbf{NOT}}\\
\cline{3-8}
I & II  & \textbf{OR} & \textbf{XOR} & \textbf{XNOR} &
\textbf{NAND} & \textbf{Input} & \textbf{Output}\\
 \hline
    0  & 0  & 0 & 0 & 52.5 & 682.2 & 0 & 52.5\\
    0  & 1  & 52.5 & 52.5 & 0 & 48.7 &1 & 0 \\
    1  & 0  & 52.5 & 52.5 & 0 & 48.7 & & \\
    1  & 1  & 686.2 & 0 & 48.7 & 0 & & \\
 \hline
\end{tabular}
\label{tab1}
\end{table}
current appears in the ring when the output signal is ON. It suggests 
an easy detection of $I_c$ by some indirect means~\cite{ind1} as the 
response involving $I_c$ will be large enough to measure. For the OFF state,
no such response will be available. In the same footing, the large $I_c$ 
induces a considerable magnetic field (few orders of milliTesla) at the 
ring center as well as in its close vicinity. The strength of magnetic 
field, associated with circular current $I_c$, can easily be estimated 
from the relation given in Eq.~\ref{bb}. Defining $R$ as the 
perpendicular distance from the centre of the benzene molecule to any 
$C$-$C$ bond, the magnetic field at the ring center can be expressed as
$B=(6\mu_0/4\pi R)I_c$. Thus, for the average circular current (say) 
$I_c=50\,\mu$A (see Table~\ref{tab1}), the induced magnetic field
(considering $R \sim 0.13\,$nm~\cite{cir2}) becomes $\sim 230\,$mT
which is too large than the required magnetic field to operate a single
spin, and it can be understood clearly from the following analysis.
It is well known that with the application of magnetic field the orientation 
of a spin can be regulated selectively. For rotating a single spin by a 
relative angle $\theta$ within a time scale $\tau$, the required magnetic 
field is~\cite{lidar}: $B=2 \hbar \theta/g \mu_B \tau$, where 
$\mu_B$ is the Bohr magneton and $g$ represents the $g$-factor. Thus,
to rotate a spin by an angle $\theta=\pi/2$, assuming the average operation
time $\tau=5$ ns, the desired magnetic field is $B \sim 7$ mT~\cite{lidar},
which is too small compared to the obtained average magnetic field 
$\sim 230\,$mT in the molecular ring. The above arguments essentially 
motivate us to recommend {\em two separate schemes for storage mechanism} 
using our molecular system. One is charge based which is our primary goal 
along with logic functions, and the other is the spin based. The ideas are 
as follows. 
For the charge based device, we mark $1$ for finite $I_c$, while it becomes 
$0$ when $I_c$ completely disappears. Since this current is non-volatile, 
the information can definitely be stored even 
when the power is off. For the other prescription i.e., spin-based, we can 
place a free magnetic site either at the center of the molecular ring or 
away (not so far) from the ring plane whose magnetic moment direction can 
be tuned by means of circular current induced magnetic field~\cite{cir2,mm1}. 
Assigning the free moment direction (not along $Z$-direction, say in the 
$X$-$Y$ plane) as $0$ for 
the case when $I_c=0$, and its direction along $Z$-axis as $1$ due to the 
interaction of magnetic moment with $B$ field for finite $I_c$, we can define 
two separate states of logic bits. As the induced magnetic field is reasonably 
high, it can easily tune a magnetic moment~\cite{lidar} which is not aligned 
initially along $Z$-direction, and we strongly believe that this prescription 
can be substantiated.

Before we end the discussion, we would like to state that the results studied
here are worked out for a typical set of parameter values. But all these results 
are equally valid for any other set of parameter values which proves the 
robustness of our analysis. 

\section{Closing Remarks}

We conclude by pointing out that a single benzene molecule (or any other 
cyclic molecule having even number of sites, though smaller rings are 
always appreciable as they can produce large circular current) is capable 
of performing different logical operations and can be programmed and 
re-programmed for suitable functional operations. Our proposal leads 
to a significant impact in designing charge based storage device along 
with functional logical operations.

\section{Acknowledgment}

MP gratefully acknowledges UGC, India (F. 2-10/2012(SA-I)) for providing her
doctoral fellowship.


\begin{thebibliography}{99}

\bibitem{mlg1} A. P. de Silva, H. Q. N. Gunaratne, and C. P. McCoy, 
A molecular photoionic AND gate based on fluorescent signalling.
Nature \textbf{364}, 42 (1993).

\bibitem{mlg2} F. M. Raymo, Digital processing and communication
with molecular switches. Adv. Mater. \textbf{14}, 401 (2002).

\bibitem{mlg3} A. P. de Silva et al., Signaling recognition events with
fluorescent sensors and switches. Chem. Rev. \textbf{97}, 1515 (1997).

\bibitem{mlg4} Y. Huang et al., Logic gates and computation from
assembled nanowire building blocks. Science \textbf{294}, 1313 (2001).

\bibitem{mlg5} S. Kou et al., Fluorescent molecular logic gates using
microfluidic devices. Angew. Chem. \textbf{120}, 886 (2008).

\bibitem{mlg6} D. Wang et al., Molecular logic gates on DNA origami
nanostructures for microRNA diagnostics. Anal. Chem. \textbf{86},
1932 (2014).

\bibitem{smnew1} M. Patra and S. K. Maiti, Logical operations using 
phenyl ring. Phys. Lett. A \textbf{382}, 420 (2018).

\bibitem{pld1} H. Dery, P. Dalal, \L{}. Cywi\'{n}ski, and L. J. Sham,
Spin-based logic in semiconductors for reconfigurable large-scale circuits.
Nature \textbf{447}, 573 (2007).

\bibitem{pld2} A. Ney, C. Pampuch, R. Koch, and K. H. Ploog,
Programmable computing with a single magnetoresistive element.
Nature \textbf{425}, 485 (2003).

\bibitem{pld3} R. Richter, L. B\"ar, J. Wecker, and G. Reiss,
Nonvolatile field programmable spin-logic for reconfigurable computing.
Appl. Phys. Lett. \textbf{80}, 1291 (2002).

\bibitem{pld4} B. Behin-Aein, D. Datta, S. Salahuddin, and S. Datta,
Proposal for an all-spin logic device with built-in memory.
Nature Nanotech. \textbf{5}, 266 (2010).

\bibitem{mr1} S. A. Wolf et al., Spintronics: A spin-based electronics
vision for the future. Science \textbf{294}, 1488 (2001).

\bibitem{stt3} Editorial. Memory with a spin. Nature Nanotech. 
\textbf{10}, 185 (2015).

\bibitem{cir1} D. Rai, O. Hod, and A. Nitzan, Circular currents in
molecular wires. J. Phys. Chem. C \textbf{114}, 20583 (2010).

\bibitem{cir2} D. Rai, O. Hod, and A. Nitzan, Magnetic fields effects on
the electronic conduction properties of molecular ring structures.
Phys. Rev. B \textbf{85}, 155440 (2012).

\bibitem{cir3} M. Patra and S. K. Maiti, Modulation of circular current
and associated magnetic field in a molecular junction: A new approach.
Sci. Rep. \textbf{7}, 43343 (2017).

\bibitem{pc1} L. P. Levy, G. Dolan, J. Dunsmuir, and H. Bouchiat,
Magnetization of mesoscopic copper rings: Evidence for persistent currents.
Phys. Rev. Lett. \textbf{64}, 2074 (1990).

\bibitem{pc2} H. F. Cheung, Y. Gefen, E. K. Riedel, and W. H. Shih,
Persistent currents in small one-dimensional metal rings.
Phys. Rev. B \textbf{37}, 6050 (1988).

\bibitem{pc3} S. K. Maiti, M. Dey, S. Sil, A. Chakrabarti, and 
S. N. Karmakar, Magneto-transport in a mesoscopic ring with Rashba and 
Dresselhaus spin-orbit interactions. Europhys. Lett. \textbf{95}, 57008 
(2011). 

\bibitem{pc4} S. K. Maiti, Determination of Rashba and Dresselhaus 
spin-orbit fields. J. Appl. Phys. \textbf{110}, 064306 (2011). 

\bibitem{pc5} M. Patra and S. K. Maiti, Unconventional low-field magnetic 
response of a diffusive ring with spin-orbit coupling. Phys. Lett. A 
\textbf{381}, 221 (2017). 

\bibitem{baer1} R. Baer and D. Neuhauser, Anti-coherence based molecular
electronics: XOR-gate response. Chem. Phys. \textbf{281}, 353 (2002).

\bibitem{baer2} R. Baer and D. Neuhauser, Phase coherent electronics: 
A molecular switch based on quantum interference. J. Am. Chem. Soc.
\textbf{124}, 4200 (2002).

\bibitem{wg1} Y. J. Xiong and X. T. Liang, Fano resonance and persistent
current of a quantum ring. Phys. Lett. A \textbf{330}, 307 (2004).

\bibitem{wg2} Y. Shi and H. Chen, Transport through an Aharonov-Casher
ring with a quantum gate. Phys. Rev. B \textbf{60}, 10949 (1999).

\bibitem{wg3} C. M. Ryu {\em et al.}, Quantum waveguide theory for triply
connected Aharonov-Bohm rings. Int. J. Mod. Phys. B \textbf{10}, 701 (1996).

\bibitem{datta} S. Datta, Electronic transport in mesoscopic systems.
Cambridge University Press, Cambridge (1997).

\bibitem{sm1} S. K. Maiti, Externally controlled local magnetic field
in a conducting mesoscopic ring coupled to a quantum wire. J. Appl.
Phys. \textbf{117}, 024306 (2015).

\bibitem{sm2} S. K. Maiti, Conformation-dependent electron transport
through a biphenyl molecule: Circular current and related issues.
Eur. Phys. J. B \textbf{86}, 296 (2013).

\bibitem{ind1} T. Heine, C. Corminboeuf, and G. Seifert, The magnetic
shielding function of molecules and Pi-electron delocalization.
Chem. Rev. \textbf{105}, 3889 (2005) and references therein.

\bibitem{mm1} K. Tagami and M. Tsukada, Current-controlled magnetism in
T-shape tape-porphyrin molecular bridges. Curr. Appl. Phys. \textbf{3},
439 (2003).

\bibitem{lidar} D. A. Lidar and J. H. Thywissen, Exponentially localized
magnetic fields for single-spin quantum logic gates. J. Appl. Phys.
\textbf{96}, 754 (2004).

\end{thebibliography}
\end{document}